\documentclass{article}
\pdfoutput=1

\usepackage{arxiv}

\usepackage[utf8]{inputenc} 
\usepackage[T1]{fontenc}    
\usepackage{hyperref}       
\usepackage{url}            
\usepackage{booktabs}       
\usepackage{amsfonts}       
\usepackage{nicefrac}       
\usepackage{microtype}      
\usepackage{graphicx}
\usepackage{gensymb}
\usepackage{upgreek}

\title{Magnetic Binary Supersaturated Solid Solutions Processed by Severe Plastic Deformation}

\author{
  Martin St\"uckler \\
  Erich Schmid Institute of Materials Science, Austrian Academy of Sciences\\
  Jahnstra{\ss}e 12, 8700 Leoben, Austria \\
  \texttt{martin.stueckler@oeaw.ac.at} 
     \And
  Heinz Krenn \\
  Institute of Physics, University of Graz\\
  Universit\"atsplatz 5, 8010 Graz, Austria 
    \And
  Reinhard Pippan\\
    Erich Schmid Institute of Materials Science, Austrian Academy of Sciences\\
  Jahnstra{\ss}e 12, 8700 Leoben, Austria 
   \And
  Lukas Weissitsch \\
  Erich Schmid Institute of Materials Science, Austrian Academy of Sciences\\
  Jahnstra{\ss}e 12, 8700 Leoben, Austria \\
   \And
   Stefan Wurster \\
     Erich Schmid Institute of Materials Science, Austrian Academy of Sciences\\
  Jahnstra{\ss}e 12, 8700 Leoben, Austria\\ 
  \And
  Andrea Bachmaier\\
  Erich Schmid Institute of Materials Science, Austrian Academy of Sciences\\
  Jahnstra{\ss}e 12, 8700 Leoben, Austria 
}

\begin{document}
\maketitle

\begin{abstract}
Samples consisting of one ferromagnetic and one diamagnetic component which are immiscible at the thermodynamic equilibrium (Co-Cu, Fe-Cu, Fe-Ag) are processed by high-pressure torsion at various compositions. The received microstructures are investigated by electron microscopy and synchrotron X-ray diffraction, showing a microstructural saturation. Results gained from microstructural investigations are correlated to magnetometry data. The Co-Cu samples show mainly ferromagnetic behavior and a decrease in coercivity with increasing Co-content. The saturation microstructure of Fe-Cu samples is found to be dual phase. Results of magnetic measurements also revealed the occurrence of two different magnetic phases in this system. For Fe-Ag, the microstructural and magnetic results indicate that no intermixing between the elemental phases takes place.
\end{abstract}


\section{Introduction}
Creating materials with tailored functional properties has been a large field of interest for many years. In the field of magnetism, for example, it was shown that the magnetic moment of $\gamma$-Fe$_2$O$_3$ can be tuned electrochemically \cite{Traunig2011}. Another approach to create materials with tunable magnetic properties is to exploit effects of magnetic dilution. Therefore, metastable materials are fabricated, consisting~of elements which exhibit large miscibility gaps in the thermodynamic equilibrium. The concept is to synthesize homogeneous structures consisting of components with different magnetic properties (e.g., ferromagnetic and diamagnetic). For instance, Chien et al. \cite{Chien1987} processed Fe-Cu samples by the vapor quenching method or Childress et al. \cite{Childress1991} produced Co-Cu samples by magnetron sputtering. Both~approaches showed that magnetic properties can be tuned, e.g., the Curie-temperature was shifted towards smaller values for high Cu-containing compositions. Such metastable solid solutions can also be processed with ball-milling \cite{Crespo1995, Ambrose1993}, leading to samples in powder form. Another route to prepare such samples is with high-pressure torsion (HPT) deformation \cite{Pippan2010a, valiev2000}, a severe plastic deformation (SPD) technique, which has the advantage that the resulting sample is already in bulk form. It was shown that binary supersaturated solid solutions can be processed at high homogeneity by HPT~\mbox{\cite{Kormout2017a, Sauvage2005}}. The~microstructure of as-deformed samples typically exhibits grain sizes in the nanocrystalline regime. First, investigations regarding magnetic tunability have been carried out on HPT-processed Co26wt.\%-Cu, showing large deviations in the saturation behaviour of the as-deformed state with respect to bulk hcp-Co \cite{Bachmaier2017}. Further investigations revealed the sensitivity of the magnetic properties, such as coercivity and remanence, on the processing parameters as well as the influence of subsequent annealing treatments as the microstructure and elemental distribution could change.

As covered by Herzer \cite{Herzer1995}, grain sizes in the nanometer regime lead to a decrease in coercivity. In~this regime, the coercivity is not controlled by domain wall motion and its hindering due to obstacles like grain boundaries. Instead, the grains' random alignment of the magnetic easy axes lead to a breakdown in coercivity \cite{Alben1978}.

In this study, supersaturated solid solutions are prepared by HPT consisting of one ferromagnetic and one diamagnetic component. The microstructure of the as-deformed samples, which are already available in bulk form, are characterized and correlated to their magnetic properties.
\section{Materials and Methods}
The investigated binary compounds consist of Co-Cu, Fe-Cu or Fe-Ag. To obtain any desired chemical composition, conventional powders are used as starting materials (Fe: MaTeck 99.9\% {$-$100$+$200~mesh,}  
Co: GoodFellow 99.9\% 50--150~$\upmu$m, Cu: AlfaAesar 99.9\% {$-$170$+$400~mesh}, Ag: AlfaAesar 99.9\% $-$500~mesh). An Ar-filled glovebox is used to store the powders and prepare the powder mixtures, which are hydrostatically compacted in Ar-atmosphere at {a nominal pressure of} 5~GPa {applied}. Pre-compacted samples are then severely deformed by HPT for 50 or 100 turns {at the same pressure} at room temperature. A detailed description of the used setup is given in Ref. \cite{Hohenwarter2009}. The~resulting samples are 8~mm in diameter and about 0.5~mm thick. Samples in the Co-Cu system are prepared at medium compositional ranges. Co37wt.\%-Cu, Co49wt.\%-Cu and Co53wt.\%-Cu samples are processed using 50 turns at the HPT, leading to shear strains of $\gamma\sim$~1500 at r~=~2~mm. Co28wt.\%-Cu and Co67wt.\%-Cu samples are processed by using 100~turns, leading to shear strains of $\gamma\sim$~3000 at r~=~2~mm. In this system, compositions with lower or higher Co-contents could not be successfully processed, due to large residual Co-particles or cracking during HPT deformation. In the Fe-Cu system, samples with low Fe-content (Fe7wt.\%-Cu, Fe14wt.\%-Cu, Fe25wt.\%-Cu) are processed using 100~turns at the HPT, leading to a shear strain of $\gamma\sim$~3000 at r~=~2~mm. Samples with higher Fe-contents fail due to shear band formation during HPT deformation. The Fe18wt.\%-Ag sample is deformed with 100~turns at the HPT ($\gamma\sim$~3000 at r~=~2~mm).

\begin{figure}
\centering
\includegraphics[width=0.58\textwidth]{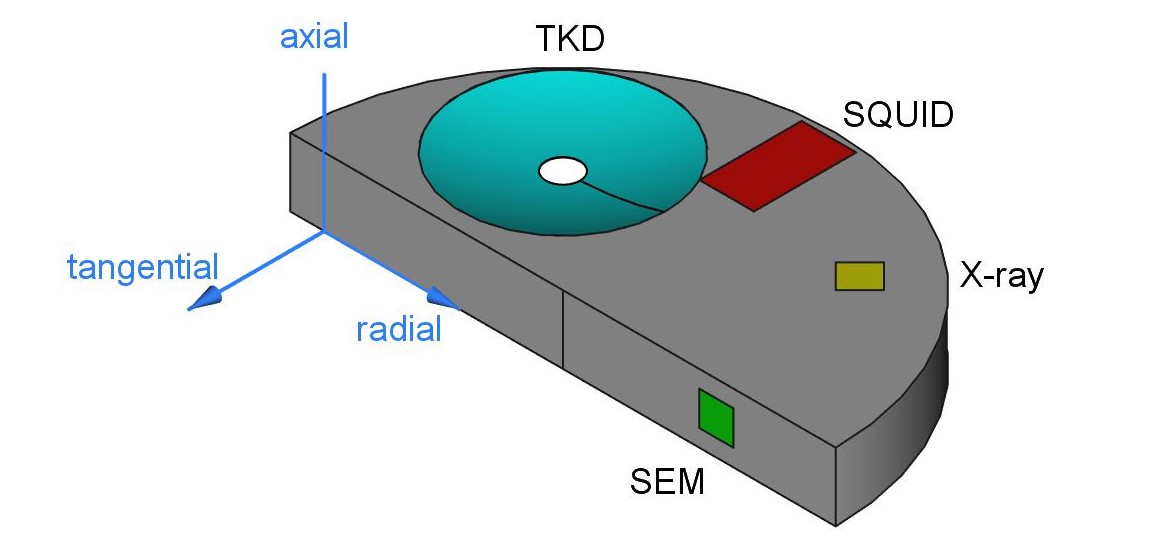}
\caption{Schematic representation of half of an {high-pressure torsion (HPT)} sample. Regions for the described measurements are highlighted. {Scanning Electron Microscopy (SEM)} images are taken in a tangential direction of the sample at r~$\geq$~2~mm, {Transmission Kikuchi Diffraction (TKD)} investigations are carried 
out in the samples axial direction. {X-ray diffraction (XRD)} measurements are carried out in transmission mode (r~$\geq$~2~mm; beam parallel to axial direction). Samples for SQUID-magnetometric measurements are cut out at r~$\geq$~2~mm.}
\label{fig:HPT_layout}
\end{figure}

Figure~\ref{fig:HPT_layout} shows a schematic diagram of an as-deformed sample and a layout of the positions where the measurements are carried out. Vickers hardness measurements are performed at half height of the sample in a tangential direction in steps of $\Delta$r~=~0.25~mm (not shown).  
Further investigations of the microstructure are carried out by scanning electron microscopy (SEM; Zeiss LEO 1525, {LEO Electron Microscopy Inc., Thornwood, USA})  
in a tangential direction of the sample. The chemical compositions of the samples, as stated above, are measured by energy dispersive X-ray spectroscopy (EDX; {Model 7426, Oxford Instruments plc, Abingdon, UK}). To investigate the grain sizes, an EBSD/TKD (Electron Back Scattering Diffraction/Transmission Kikuchi Diffraction) system ({Bruker Nano GmbH, Berlin, Germany}) attached to the SEM was used. For TKD data analysis, the {manufacturer's} software Esprit version 2.1 was utilized. Additional microstructural investigations are carried out with synchrotron X-ray diffraction measurements in transmission mode ({Petra III: P07 synchrotron facility at Deutsches Elektronen-Synchrotron DESY,
 Hamburg, Germany}; Beam Energy: 100~keV; Beam Size: 0.2 $\times$ 0.2~mm$^2$). Diffraction patterns are recorded in the axial orientation at r~$\geq$~2~mm. 

Magnetic properties are measured in a SQUID-Magnetometer (Quantum Design MPMS-XL-7, {Quantum Design, Inc., San Diego, USA}) operated with the {manufacturer's} software MPMS MultiVu Application (version 1.54). The applied magnetic field points in the axial orientation of the sample. Magnetic hysteresis are measured at 300~K in magnetic fields up to 70~kOe. Zero Field Cooling (ZFC) and Field Cooling (FC) measurements are recorded between 5~K and 300~K at 50~Oe.  

\section{Results and Disussion}
\vspace{-6pt}
\subsection{Microstructure}
Vickers hardness measurements are carried out as the measurements are very sensitive to small changes in the microstructure. Hardness values plotted versus shear strain showed a saturation behavior starting at r~$\geq$~2~mm, indicating the as-deformed microstructures reached a steady state above this position.

\begin{figure}
\centering
\includegraphics[width=0.325\textwidth]{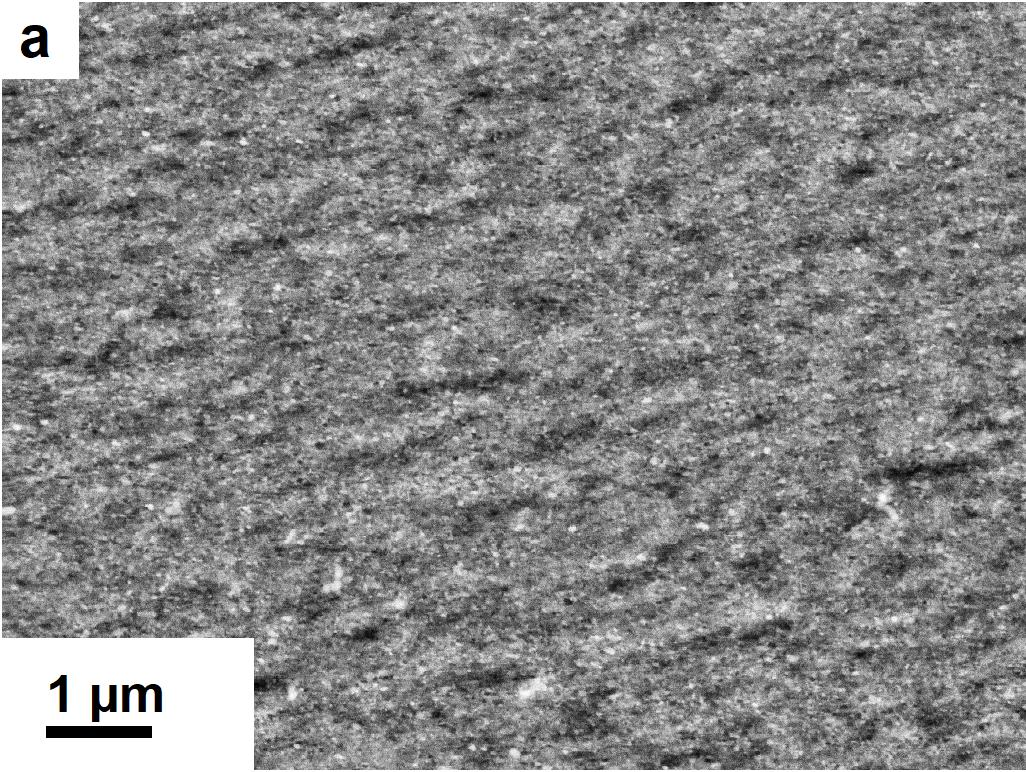}
\includegraphics[width=0.325\textwidth]{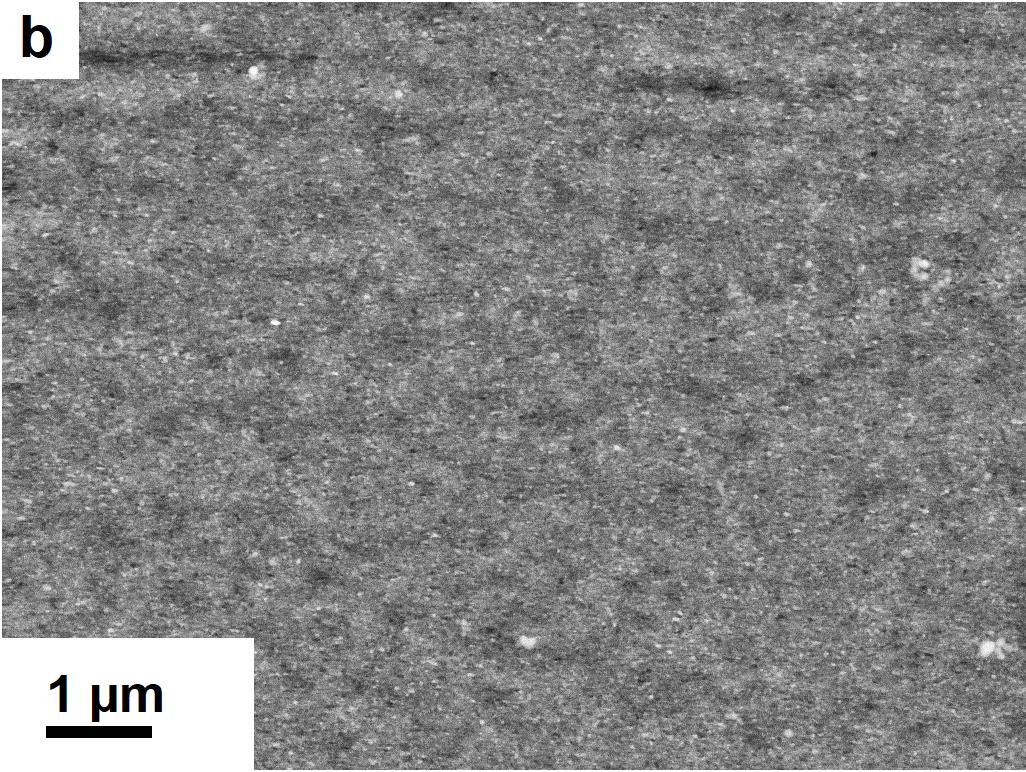}
\includegraphics[width=0.325\textwidth]{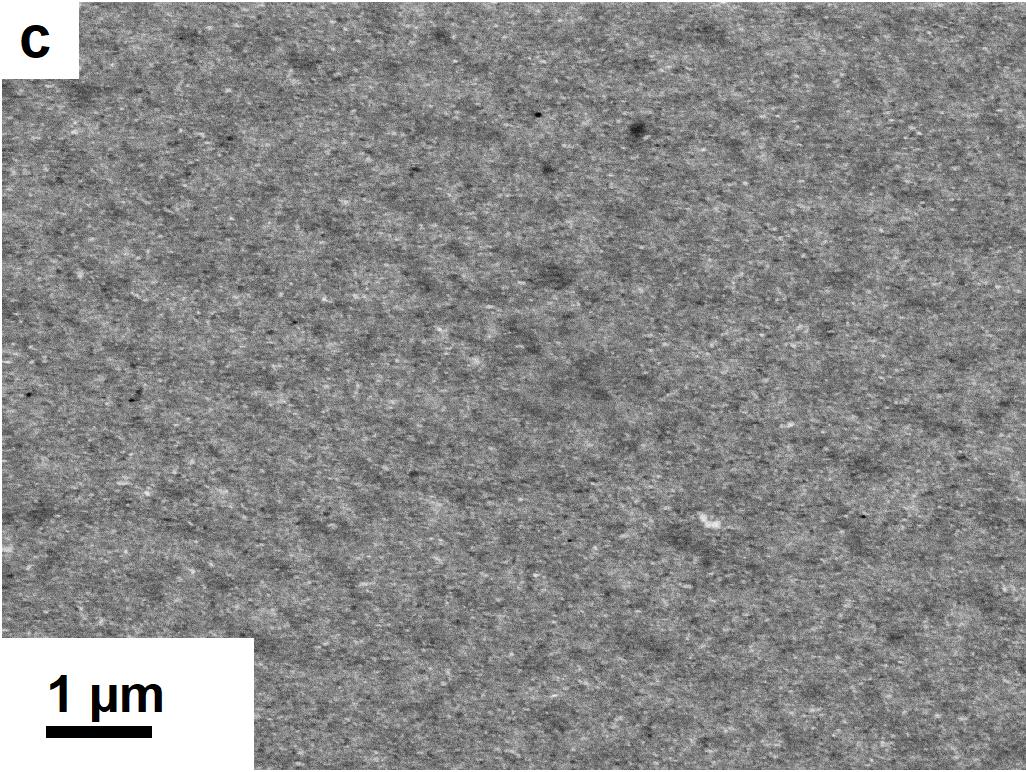}
\includegraphics[width=0.325\textwidth]{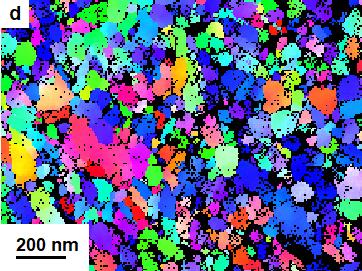}
\includegraphics[width=0.325\textwidth]{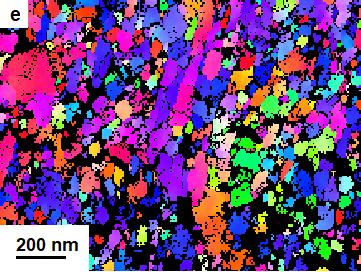}
\includegraphics[width=0.325\textwidth]{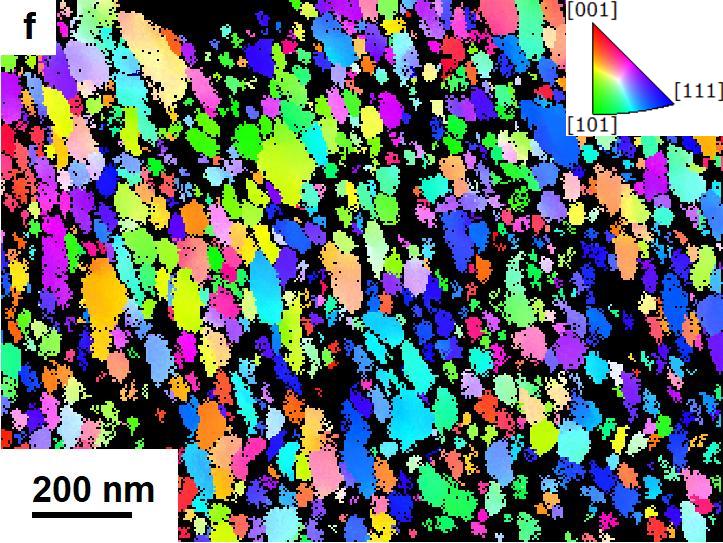}
\caption{Backscattered Electrons (BSE) micrographs of as-deformed Co-Cu samples at r~$\geq$~2~mm in tangential direction. Images of  Co28wt.\%-Cu (\textbf{a}); Co49wt.\%-Cu (\textbf{b}); and Co67wt.\%-Cu (\textbf{c}) show homogeneous microstructures. In (\textbf{d}--\textbf{f}), the corresponding TKD images, taken along the axial direction, are shown. High Co-containing compositions show smaller grain sizes.}
\label{fig:SEM_Co}
\end{figure}

\begin{figure}
\centering
\includegraphics[width=0.325\textwidth]{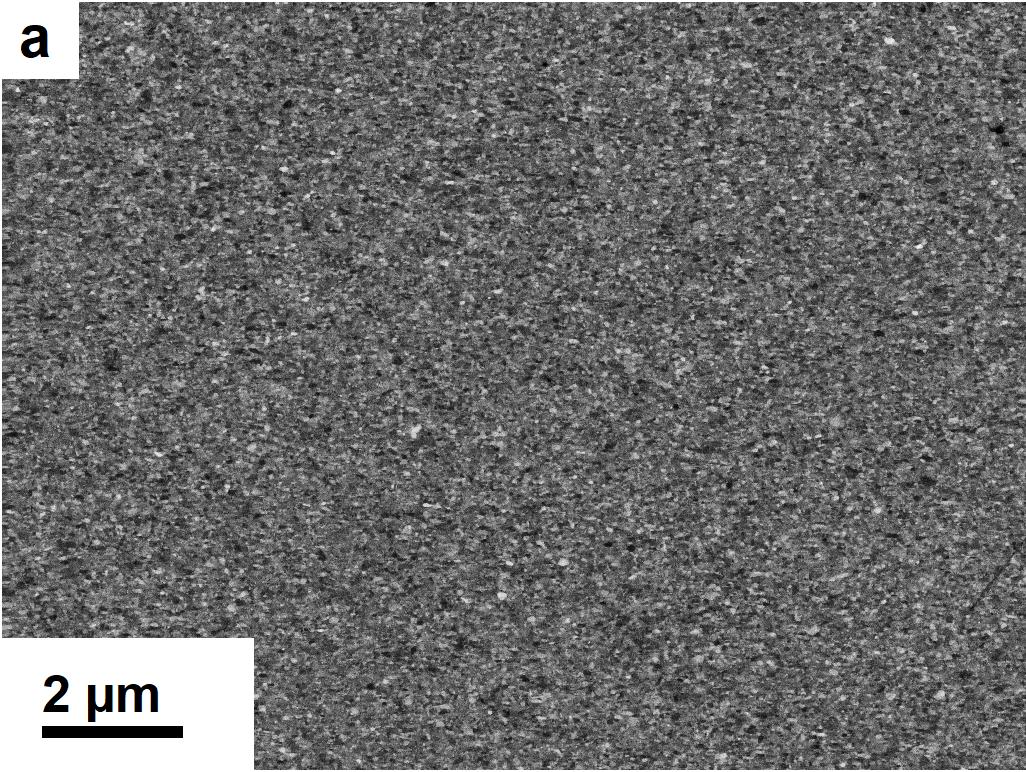}
\includegraphics[width=0.325\textwidth]{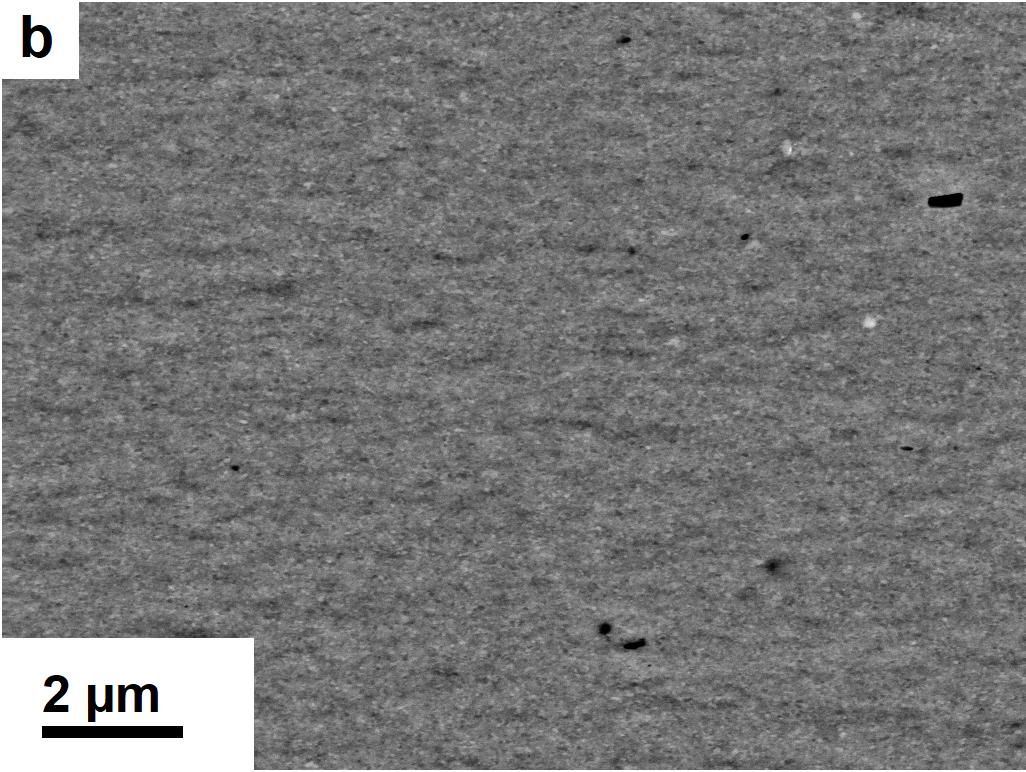}
\includegraphics[width=0.325\textwidth]{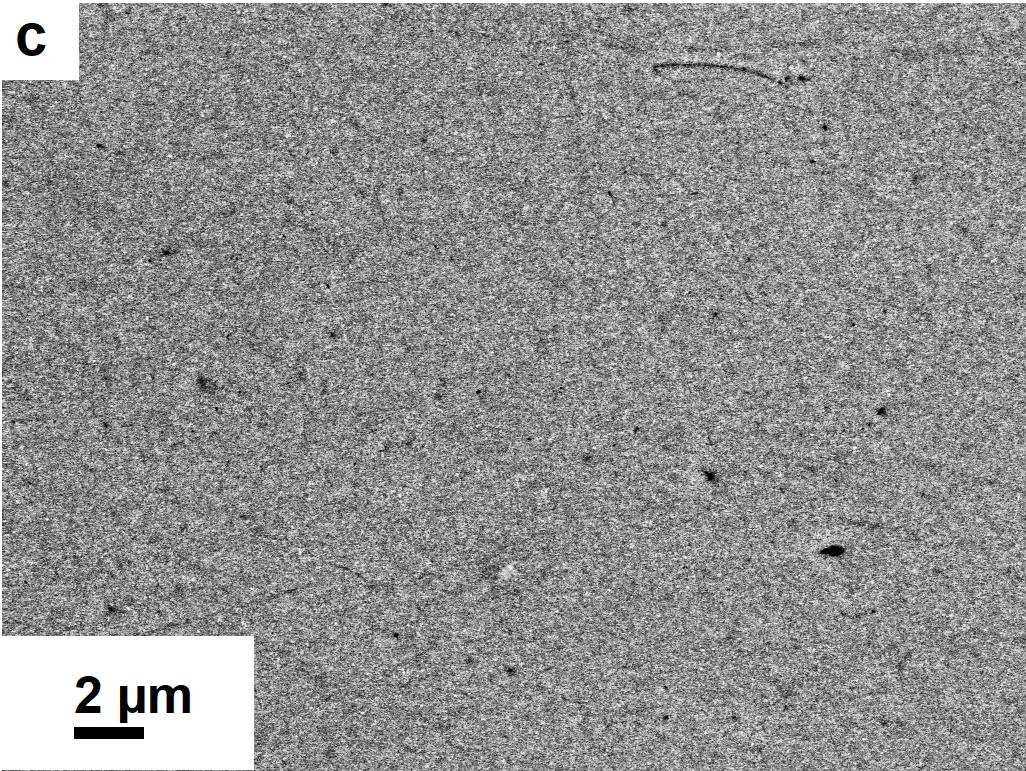}
\caption{BSE micrographs of as-deformed Fe-based samples at r~$\geq$~2~mm along the tangential direction. The microstructure of Fe7wt.\%-Cu (\textbf{a}) appears homogenous. Fe25wt.\%-Cu (\textbf{b}) exhibits some remaining particles embedded in a highly homogeneous matrix. Fe18wt.\%-Ag (\textbf{c}) shows high contrast variations indicating less Fe dissolved in the Ag-matrix.}
\label{fig:SEM_Fe}
\end{figure}

\begin{figure}
\centering
\includegraphics[width=0.49\textwidth]{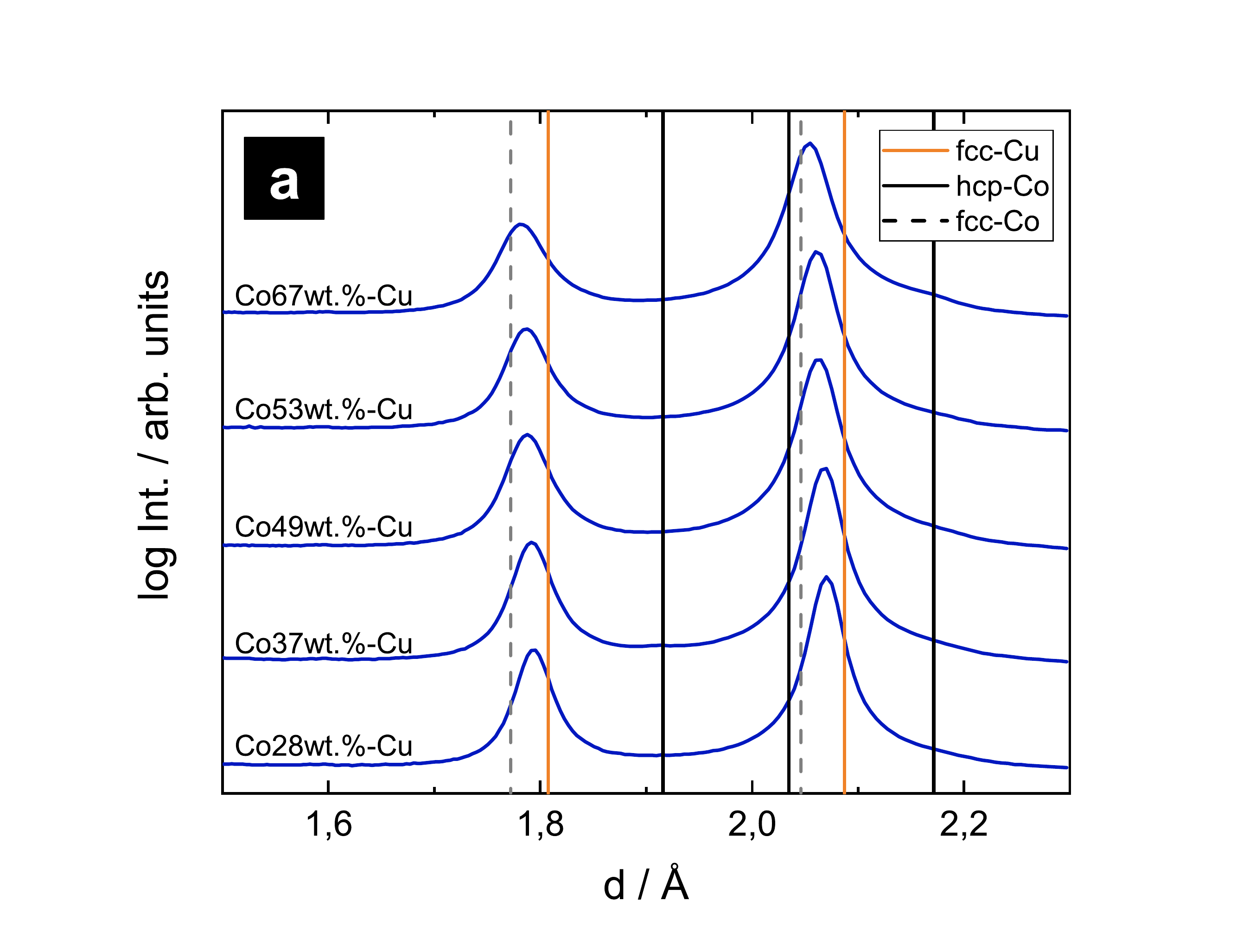}
\includegraphics[width=0.49\textwidth]{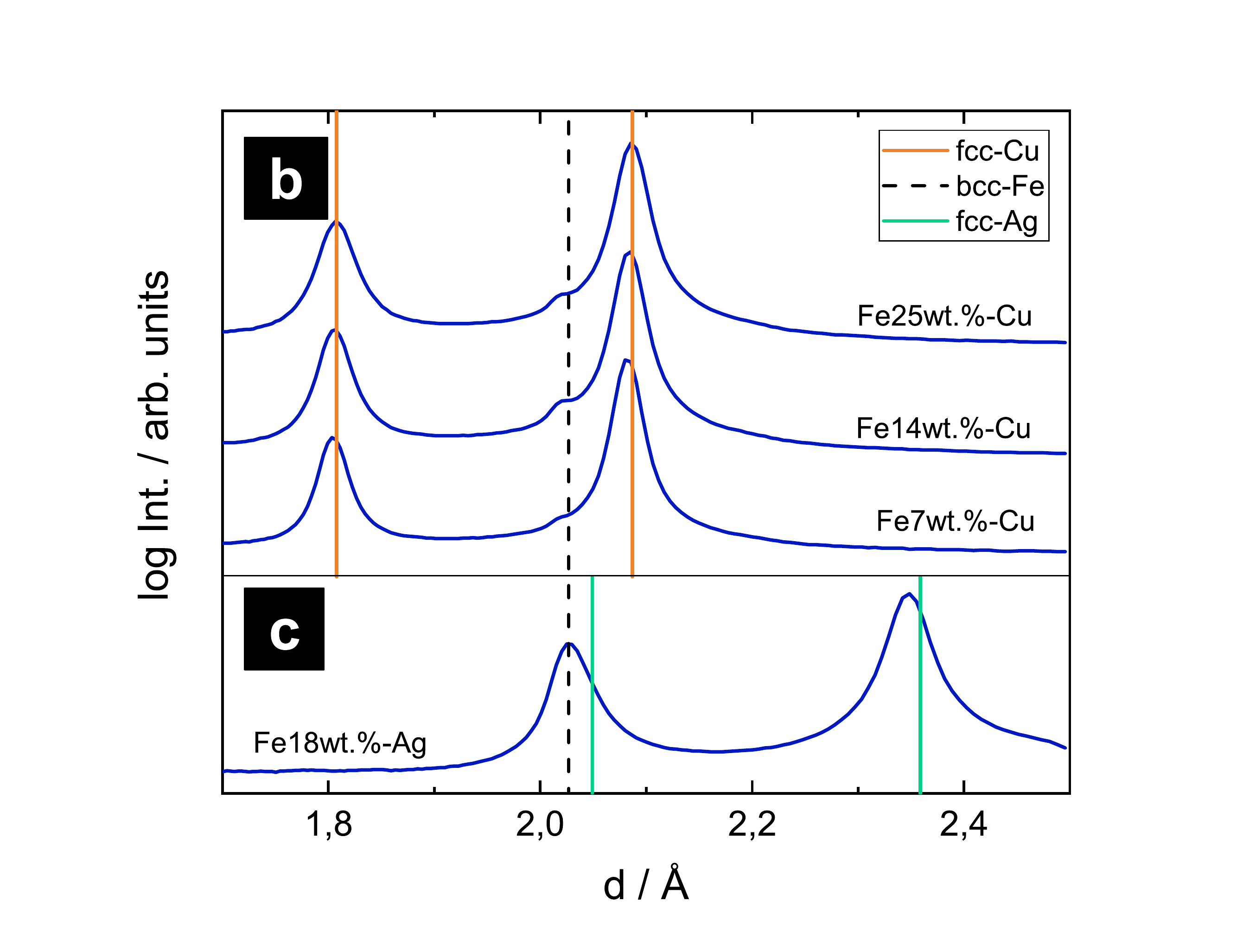}
\caption{Synchrotron {XRD} patterns of as-deformed samples. In the patterns of Co-Cu samples (\textbf{a}), only one set of fcc-peaks remains; (\textbf{b}) XRD patterns of Fe-Cu samples show pronounced fcc-Cu peaks and weak bcc-Fe peaks. In the XRD pattern of Fe18wt.\%-Ag, (\textbf{c}) bcc-Fe and fcc-Ag peaks overlap.}
\label{fig:XRD}
\end{figure}

Figure~\ref{fig:SEM_Co}{a}--{c} show SEM images of as-deformed Co-Cu samples, taken in a tangential direction of the sample at r~$\geq$~2 mm (back-scatter mode). The micrographs indicate a homogeneous phase and show very small grain sizes in the nanocrystalline regime. Figure~\ref{fig:SEM_Co}{d}--{f} show TKD orientation maps of similar samples, taken in an axial direction. At least three EBSD-scans per specimen were jointly analyzed to evaluate the area weighted grain sizes. Taking only high angle boundaries into account leads to the following results: 100~nm, 78~nm and 77~nm for Co28wt.\%-Cu, Co49wt.\%-Cu, Co67wt.\%-Cu,~respectively.
SEM images of Fe-Cu with low (7 wt.\%) and high (25 wt.\%) Fe-contents are shown in Figure~\ref{fig:SEM_Fe}{a},{b}. A few remaining particles are visible in the micrograph of Fe25wt.\%-Cu. In Figure~\ref{fig:SEM_Fe}{c}, the SEM image of Fe18wt.\%-Ag is displayed. The deformed microstructure exhibits many remaining dark particles at various sizes.
Statistically significant information of each sample's constituting phases are revealed by synchrotron diffraction measurements, as shown in Figure~\ref{fig:XRD}. In the diffraction patterns of the Co-Cu samples (Figure~\ref{fig:XRD}{a}), {only very weak}  occurence of hcp-Co {may persist}. The received patterns consist {mainly}  of fcc-peaks, showing that Co undergoes a phase transformation from hcp to fcc during HPT as reported in previous studies \cite{Bachmaier2015, Edalati2013, sort2003microstructural}. The deviations of the fcc-Cu peaks can be explained with the change of the chemical composition of the samples, summarizing that the received microstructures {is mainly} fcc, either rich in Cu or Co.
Diffraction patterns of Fe-Cu (Figure~\ref{fig:XRD}{b}) samples exhibit pronounced peaks of fcc-Cu, but also weak bcc-Fe peaks can be identified. These results are in agreement with SEM images as described above, which also show a few remaining particles and leads to the conclusion that a dual phase structure is present. The diffraction pattern of the Fe18wt.\%-Ag sample is shown in Figure~\ref{fig:XRD}{c}. The~occurring phases in the pattern can not be clearly identified due to an overlap in the bcc-Fe and fcc-Ag peaks. These peaks are very broad, indicating a remaining dual phase structure.
\subsection{Magnetism}   
\begin{figure}
\centering
\includegraphics[width=0.49\textwidth]{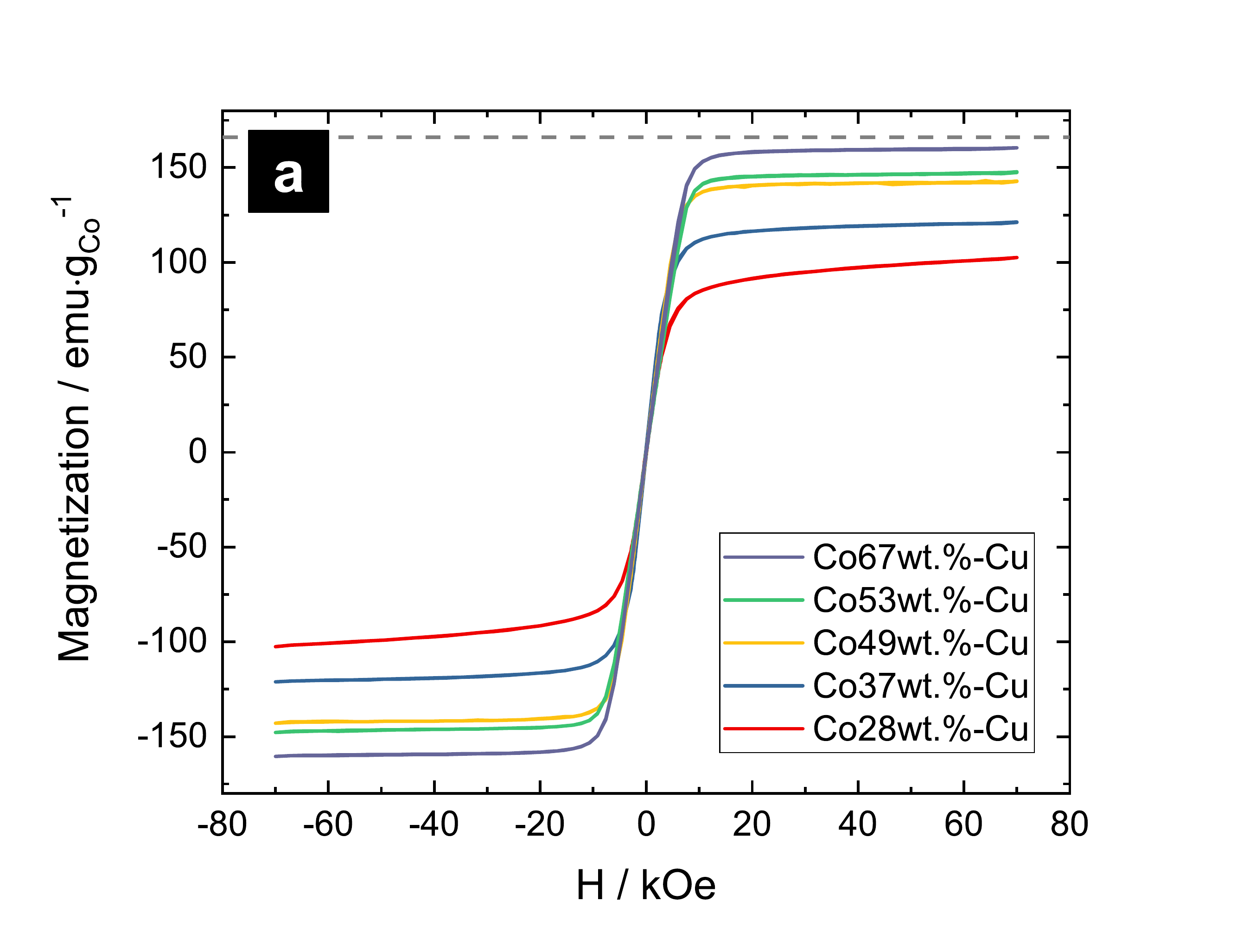}
\includegraphics[width=0.49\textwidth]{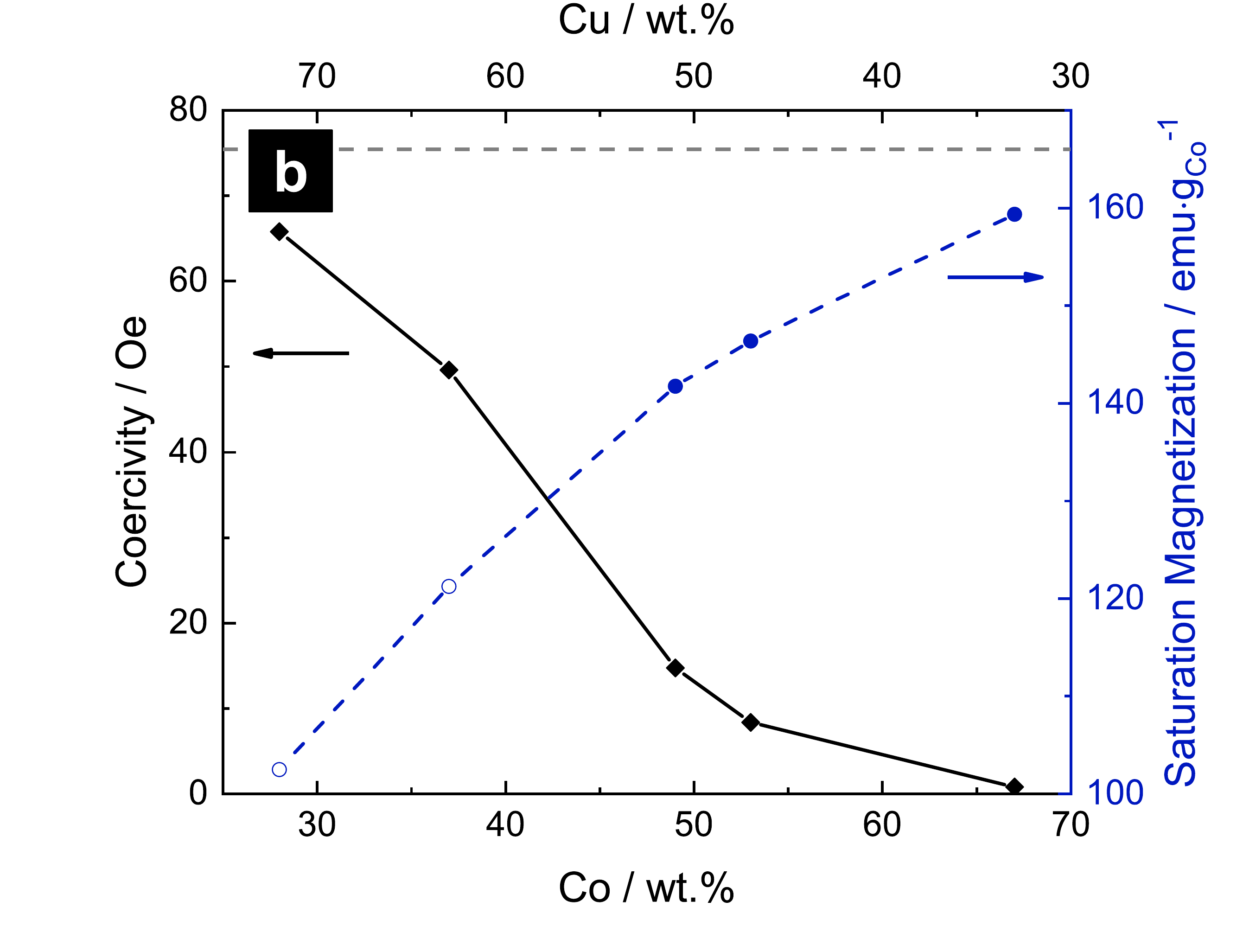}
\caption{(\textbf{a}) magnetization versus applied field for Co-Cu samples measured at 300 K; (\textbf{b}) coercivity and saturation magnetization are plotted versus Co-content. The grey dotted lines in both figures indicate the magnetic moment of fcc-Co (166~emu$\cdot$g$^{-1}$). Full saturation is not achieved for Co contents $\leq$~40 wt.\% in a field of 70~kOe (open symbols in (\textbf{b})).}
\label{fig:CoCu_hyst}
\end{figure}

Figure~\ref{fig:CoCu_hyst}{a} shows the hysteresis of Co-Cu samples measured at 300~K. The magnetic moment per gram of cobalt is plotted versus the applied field. The three samples with the highest Co-content (Co49wt.\%-Cu, Co53wt.\%-Cu, Co67wt.\%-Cu) saturate easily, indicating a pronounced ferromagnetic ordering. The saturation behavior of the samples with lower Co-content, Co28wt.\%-Cu and Co37wt.\%-Cu illustrate a slight increase of the magnetization with increasing magnetic field. Saturation is not completed even at the highest applied field of 70~kOe, which indicates a paramagnetic contribution and, therefore, a partial breakdown in the long-range ordering. In Figure \ref{fig:CoCu_hyst}{b}, the~magnetic moment per gram of Co in saturation is plotted versus the Co-content. Therefore, the mean of the magnetic moment was calculated at fields between 20~kOe and 70~kOe (filled symbols). For the samples with low Co-content, Co28wt.\%-Cu and Co37wt.\%-Cu, the magnetization at 70~kOe is plotted (open symbols). It can be seen that the magnetic moment of Co increases with increasing Co-content and approaches the magnetic moment of bulk fcc-Co (166~emu$\cdot$g$^{-1}$ \cite{LandoltBornstein1986}). This result is in accordance with findings on magnetron sputtered Co-Cu, but slightly shifted towards higher magnetic moments~\cite{Childress1991}. The coercivity is evaluated with a least-squares fit of the hysteresis two halves between $-$800~Oe and 800~Oe and plotted in Figure~\ref{fig:CoCu_hyst}{b}. The coercivity decreases with increasing Co-content, reaching from 66~Oe down to 0.8~Oe.

From the TKD analysis, it was shown that high Co-containing compositions exhibit smaller grain sizes than low Co-containing compositions. Comparing the coercivity of the samples with their grain sizes, it can be seen that the coercivity decreases with decreasing grain size, entering the regime of random anisotropy due to exchange coupled nanograins. Apart from the decrease in grain size, the~various chemical compositions may also lead to deviations of the micromagnetic properties.  

\begin{figure}
\centering
\includegraphics[width=0.49\textwidth]{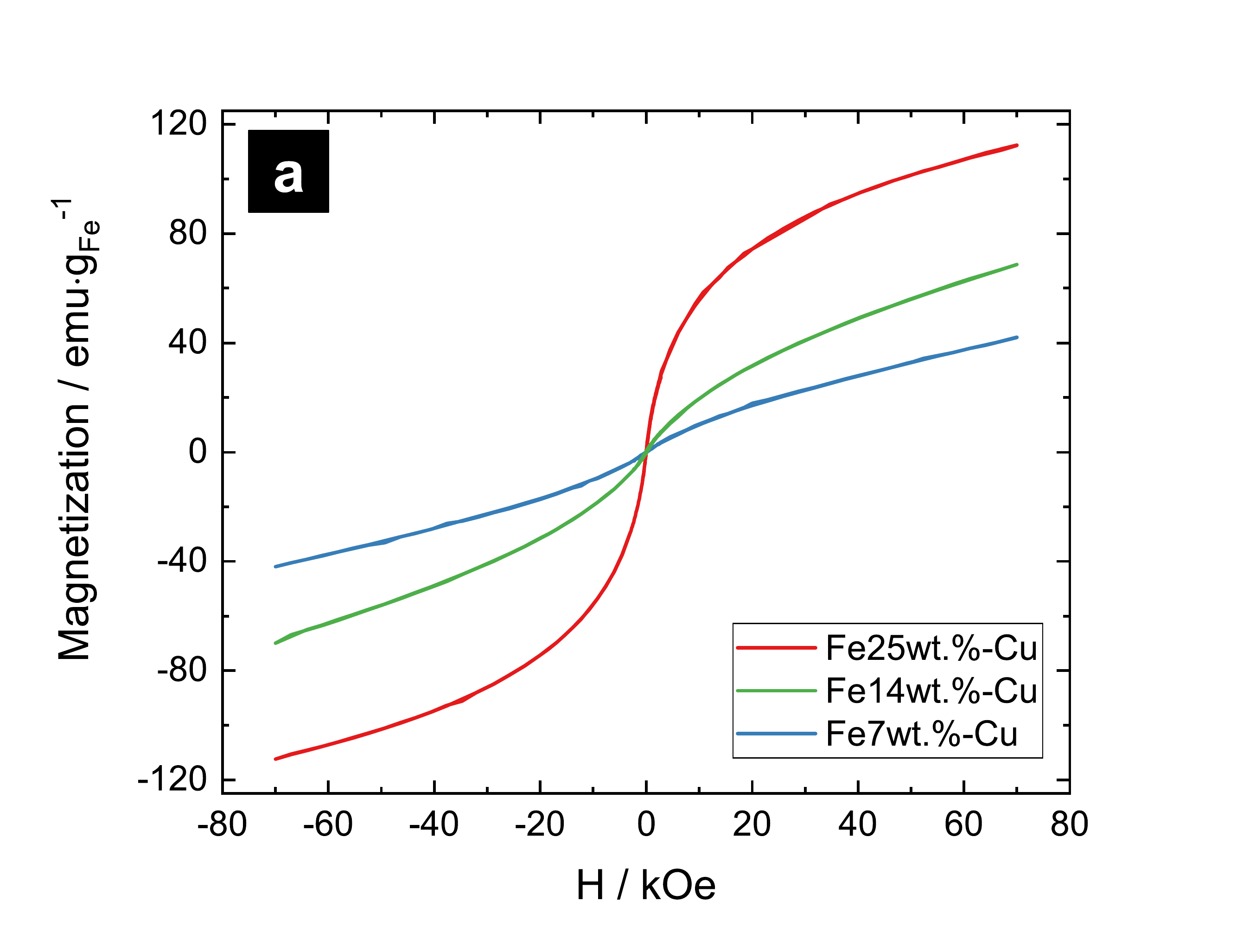}
\includegraphics[width=0.3\textwidth]{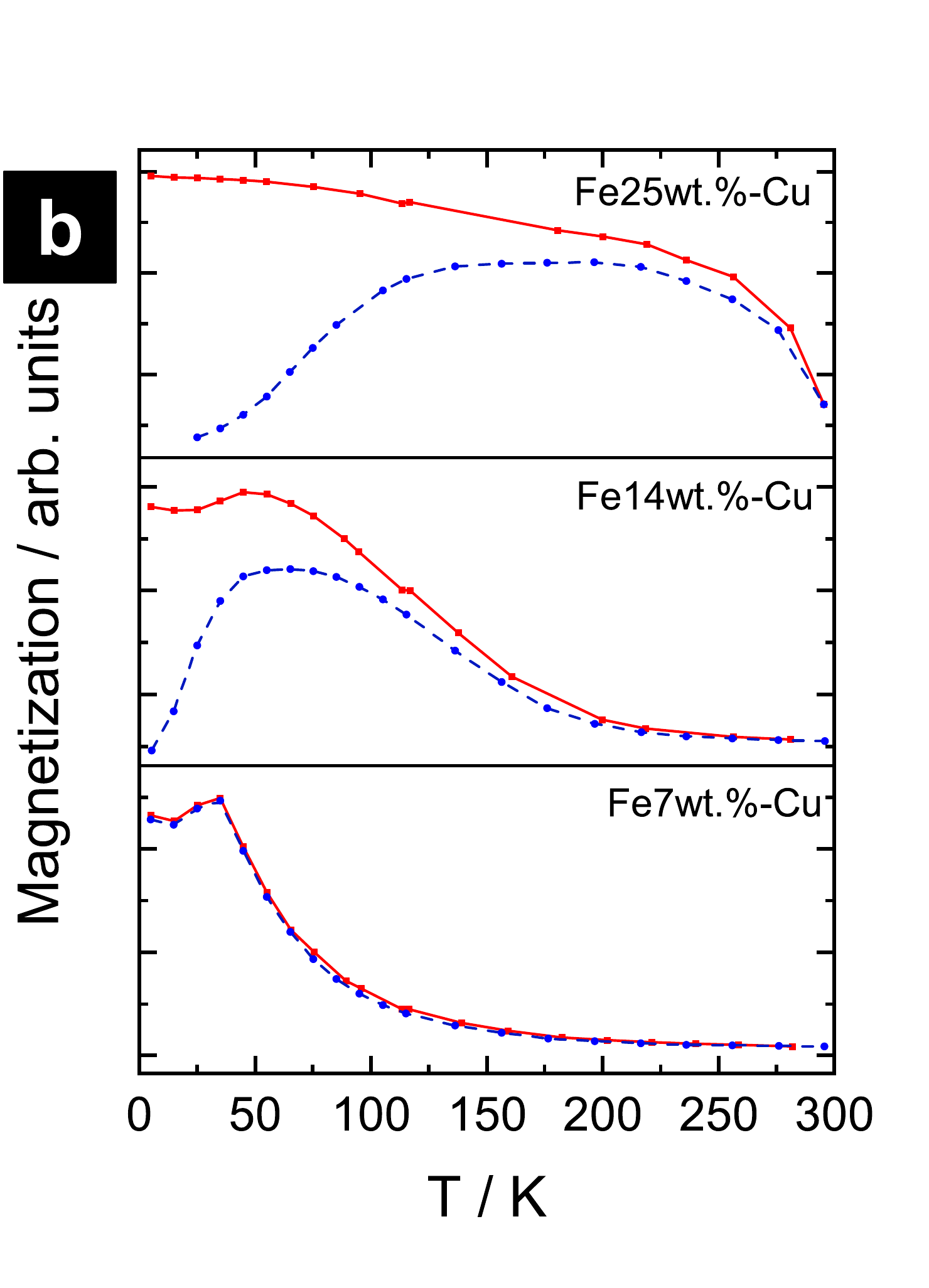}
\caption{(\textbf{a}) magnetization versus applied field for Fe-Cu samples measured at 300~K. Saturation is not achieved, even at the highest applied field; (\textbf{b}) ZFC-FC for Fe-Cu samples at 50~Oe. Splitting is not pronounced for Fe7wt.\%-Cu. For Fe7wt.\%-Cu and Fe14wt.\%-Cu a cusp at the FC-curve is found.}
\label{fig:FeCu_hyst}
\end{figure}

\begin{figure}
\centering
\includegraphics[width=0.49\textwidth]{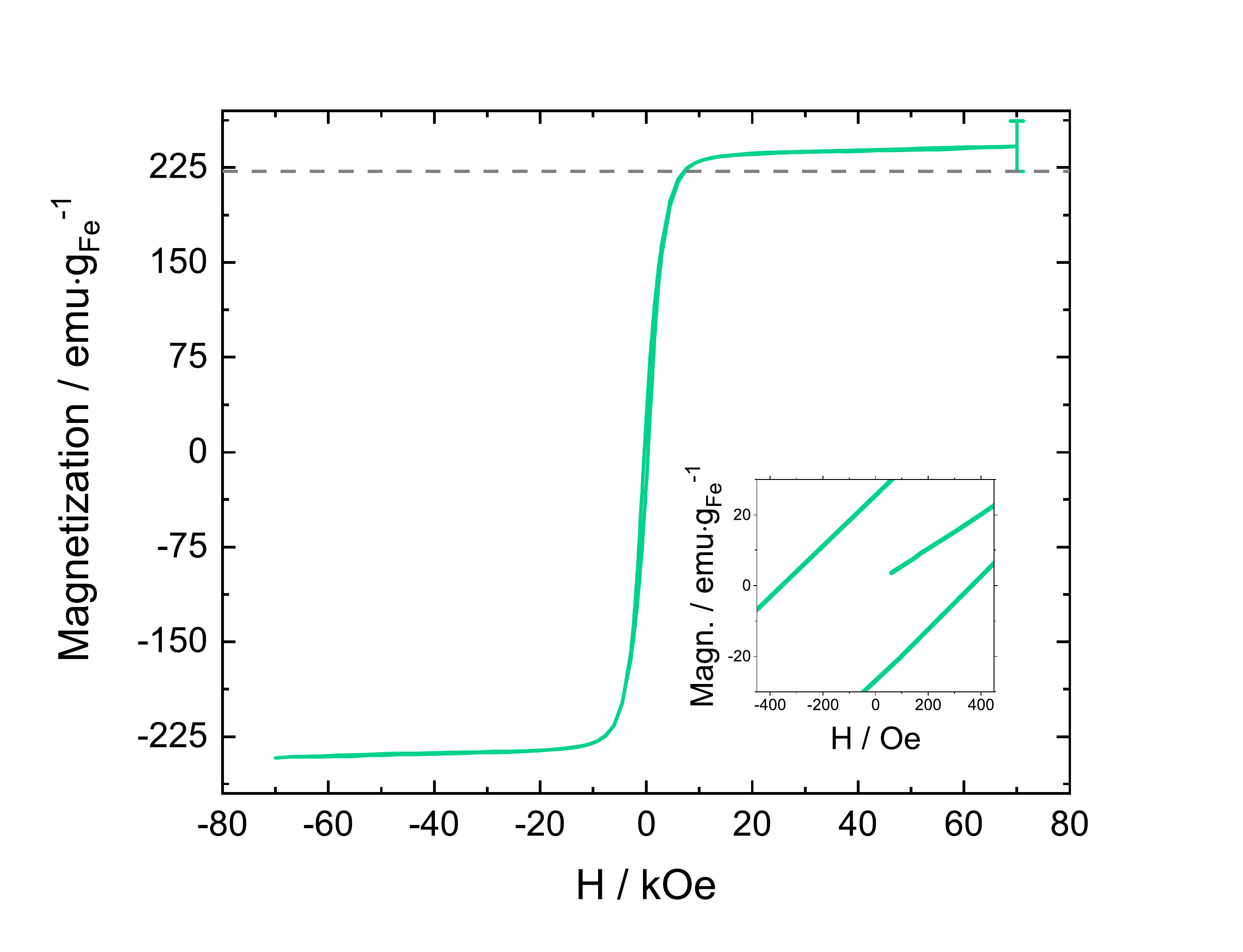}
\caption{Magnetization versus applied field for Fe18wt.\%-Ag measured at 300~K. The saturation magnetization is 240~$\pm$~20~emu$\cdot$g$^{-1}$, the coercivity is 400~Oe.}
\label{fig:FeAg_hyst}
\end{figure}

In Figure~\ref{fig:FeCu_hyst}{a}, the hysteresis of the Fe-Cu samples are shown. The magnetic moment per gram of Fe is plotted versus applied field. The hysteresis of the Fe-Cu samples show a very pronounced paramagnetic behavior at high applied fields. At low fields, a steep increase in magnetization occurs, indicating the presence of long-range magnetic ordering and, therefore, a ferromagnetic phase as expected from SEM and sychrotron XRD measurements. This contribution increases with increasing Fe-content. The magnetization, even at the highest applied field, is far away from the value of bulk bcc-Fe at 222~emu$\cdot$g$^{-1}$ for all investigated compositions. This can be explained by the disorder of surface spins on genuine Fe clusters formed by HPT deformation. Further magnetic characterization is carried out by performing temperature dependent measurements at low magnetic field (50~Oe). Results~of ZFC-FC measurements on the Fe-Cu samples are shown in Figure~\ref{fig:FeCu_hyst}{b}. In the ZFC-FC measurement of Fe25wt.\%-Cu and Fe14wt.\%-Cu, superparamagnetic blocking peaks below room temperature can be identified, indicating the presence of remaining Fe clusters. Splitting of the ZFC-FC-curves is not pronounced for the Fe7wt.\%-Cu sample, but Fe7wt.\%-Cu and Fe14wt.\%-Cu show a cusp in the FC curve at very low temperatures. Its origin may be found in spin-glass behaviour, as reported in Refs. \cite{Adachi1986, Franz1973, Vedyaev1982}. This phase transition is expected to apply for Fe-contents below 20 at.\% at about 50~K \cite{Chien1987}, which leads to the assumption that Fe is not only clustered, but is also partially diluted in Cu. As the expected magnetic effects for Fe-Cu samples are highly sensitive to the elemental distributions, further microstructural investigations need to be carried out at atomic scales (e.g., by~Atom Probe Tomography) to get more in-depth information.
The hysteresis of the Fe18wt.\%-Ag sample is shown in Figure~\ref{fig:FeAg_hyst}. Any deviation between the saturation magnetization and the magnetic moment of bulk bcc-Fe is in the range of the chemical composition uncertainty, leading to the conclusion that no distortion of the Fe magnetic moment is observed. The coercivity, evaluated as described above, is 400~Oe. These results again indicate that no intermixing between Ag and Fe takes place during HPT and a dual-phase structure remains. It can be concluded that, for the Fe18wt.\%-Ag sample, only grain refinement takes place.


\section{Conclusions}
Binary solid solutions are processed by HPT. Three different systems, immiscible at the thermodynamic equilibrium and consisting of one ferromagnetic and one diamagnetic component, are investigated (Co-Cu, Fe-Cu, Fe-Ag). Correlating microstructural and magnetic data lead to the following results: for Co-Cu samples, fcc-structures can be processed in the medium composition range. Higher Co-containing compositions show the smallest grain sizes as well as the lowest coercivity, demonstrating that the coercivity can be tuned by varying the chemical composition. Co-Cu HPT samples are near the crossover of free-domain motion and exchange-coupling of ultra-small grains exhibiting randomization of anisotropy, with nearly identical initial susceptibilites independent of Co-composition. On the other hand, Fe-Cu samples deliver phase separation of Fe-grains with monodomain magnetism obeying Stoner--Wolfarth behavior with strongly composition dependent slopes of hysteresis curves and a typical magnetic blocking effect. For Fe-concentrations below 20~wt.\%, a partial dissolution of Fe in Cu is expected with spin-glass behavior (to be probed by frequency dependent AC 
susceptibility measurements). For the Fe-Ag sample, magnetic measurements indicate only grain refinement, but no intermixing of the elemental phases takes place.


\vspace{6pt} 



\section*{Author Contributions}
Conceptualization, M.S., R.P. and A.B.; Methodology, M.S., S.W. and A.B.; Validation, M.S., H.K., R.P., L.W., S.W. and A.B.; Formal Analysis, M.S., Investigation, M.S. and S.W.; Resources, H.K.; Data Curation, M.S.; Writing---Original Draft Preparation, M.S.; Writing---Reviewing and Editing H.K., R.P., L.W., S.W. and A.B.; Visualization, M.S.; Supervision, R.P and A.B.; Funding Acquisition, A.B.

\section*{Funding}
This project has received funding from the European Research Council (ERC) under the European Union’s Horizon 2020 research and innovation programme (Grant No. 757333).

\section*{Acknowledgements} The measurements leading to these results have been performed at PETRA III: P07 at DESY Hamburg (Germany), a member of the Helmholtz Association (HGF). We gratefully acknowledge the assistance by Norbert Schell. The authors thank Karoline Kormout, Sandra Ebner, Christina Hofer and Stefan Zeiler for their help with the synchrotron measurements. 

\section*{Conflicts of Interest}
The authors declare no conflict of interest.

\section*{Abbreviations}
The following abbreviations are used in this manuscript:\\

\noindent 
\begin{tabular}{@{}ll}
BSE & Backscattered Electrons\\
EBSD & Electron Backscatter Diffraction\\
EDX & Energy Dispersive X-Ray Spectroscopy\\
FC  & Field Cooling\\
HPT & High-Pressure Torsion\\
SEM & Scanning Electron Microscopy\\
SPD & Severe Plastic Deformation\\
SQUID & Superconducting Quantum Interference Device\\
TKD & Transmission Kikuchi Diffraction\\
ZFC & Zero Field Cooling\\
\end{tabular}

\bibliography{magnbin}

\begin{thebibliography}{10}

\bibitem{Traunig2011}
Thomas Trau{\ss}nig, Stefan Topolovec, Kashif Nadeem, Doroth{\'{e}}e {Vinga
  Szab{\'{o}}}, Heinz Krenn, and Roland W{\"{u}}rschum.
\newblock {Magnetization of Fe-oxide based nanocomposite tuned by surface
  charging}.
\newblock {\em PHYS STATUS SOLIDI}, 5(4):150--152, 2011.

\bibitem{Chien1987}
C~L Chien, S~H Liou, and M~A Gatzke.
\newblock {Magnetic Percolation in new crystalline fcc Fe-Cu Alloys}.
\newblock {\em MRS Proceedings}, 80:395--400, 1987.

\bibitem{Childress1991}
J~R Childress and C~L Chien.
\newblock {Reentrant magnetic behavior in fcc Co-Cu alloys}.
\newblock {\em PHYS REV B}, 43(10):8089--8093, 1991.

\bibitem{Crespo1995}
P~Crespo, I~Navarro, A~Hemando, P~Rodr{\'{i}}guez, A~{Garc{\'{i}}a Escorial},
  J~M Barandiar{\'{a}}n, O~Drbohlav, and A~R Yavad.
\newblock {Magnetic and structural properties of as-milled and heat-treated
  bcc-Fe$_{70}$Cu$_{30}$ alloy}.
\newblock {\em J MAGN MAGN MATER}, 150:409--416, 1995.

\bibitem{Ambrose1993}
T~Ambrose, A~Gavrin, and C~L Chien.
\newblock {Magnetic propeties of metastable fcc Fe-Cu alloys prepared by high
  energy ball milling}.
\newblock {\em J MAGN MAGN MATER}, 124:15--19, 1993.

\bibitem{Pippan2010a}
R.~Pippan, S.~Scheriau, A.~Taylor, M.~Hafok, A.~Hohenwarter, and A.~Bachmaier.
\newblock {Saturation of Fragmentation During Severe Plastic Deformation}.
\newblock {\em ANN REV MATER RES}, 40(1):319--343, 2010.

\bibitem{valiev2000}
Ruslan~Zafarovich Valiev, Rinat~K Islamgaliev, and Igor~V Alexandrov.
\newblock Bulk nanostructured materials from severe plastic deformation.
\newblock {\em PROG MATER SCI}, 45(2):103--189, 2000.

\bibitem{Kormout2017a}
K.~S. Kormout, R.~Pippan, and A.~Bachmaier.
\newblock {Deformation-Induced Supersaturation in Immiscible Material Systems
  during High-Pressure Torsion}.
\newblock {\em ADV ENG MATER}, 19(4):1--19, 2017.

\bibitem{Sauvage2005}
X.~Sauvage, F.~Wetscher, and P.~Pareige.
\newblock {Mechanical alloying of Cu and Fe induced by severe plastic
  deformation of a Cu-Fe composite}.
\newblock {\em ACTA MATER}, 53(7):2127--2135, 2005.

\bibitem{Bachmaier2017}
A~Bachmaier, H~Krenn, P~Knoll, H~Aboulfadl, and R~Pippan.
\newblock {Tailoring the magnetic properties of nanocrystalline Cu-Co alloys
  prepared by high-pressure torsion and isothermal annealing}.
\newblock {\em J ALLOY COMPD}, 725:744--749, 2017.

\bibitem{Herzer1995}
G~Herzer.
\newblock {Soft-magnetic nanocrystalline materials}.
\newblock {\em SCRIPTA METALL MATER}, 33(10-11):1741--1756, 1995.

\bibitem{Alben1978}
R~Alben, J~J Becker, and M~C Chi.
\newblock {Random anisotropy in amorphous ferromagnets}.
\newblock {\em J APPL PHYS}, 49(3):1653--1658, 1978.

\bibitem{Hohenwarter2009}
A.~Hohenwarter, A.~Bachmaier, B.~Gludovatz, S.~Scheriau, and R.~Pippan.
\newblock {Technical parameters affecting grain refinement by high pressure
  torsion}.
\newblock {\em INT J MATER RES}, 100(12):1653--1661, 2009.

\bibitem{Bachmaier2015}
A.~Bachmaier, M.~Pfaff, M.~Stolpe, H.~Aboulfadl, and C.~Motz.
\newblock {Phase separation of a supersaturated nanocrystalline Cu-Co alloy and
  its influence on thermal stability}.
\newblock {\em ACTA MATER}, 96:269--283, 2015.

\bibitem{Edalati2013}
Kaveh Edalati, Shoichi Toh, Makoto Arita, Masashi Watanabe, and Zenji Horita.
\newblock {High-pressure torsion of pure cobalt: Hcp-fcc phase transformations
  and twinning during severe plastic deformation}.
\newblock {\em APPL PHYS LETT}, 102(18):1--5, 2013.

\bibitem{sort2003microstructural}
J~Sort, A~Zhilyaev, M~Zielinska, J~Nogu{\'e}s, S~Surinach, J~Thibault, and
  MD~Bar{\'o}.
\newblock Microstructural effects and large microhardness in cobalt processed
  by high pressure torsion consolidation of ball milled powders.
\newblock {\em ACTA MATER}, 51(20):6385--6393, 2003.

\bibitem{LandoltBornstein1986}
M.~B. Stearns.
\newblock 1.1.2.4 spontaneous magnetization, magnetic moments and high-field
  susceptibility: Datasheet from landolt-b{\"o}rnstein - group iii condensed
  matter {\textperiodcentered} volume 19a: ``3d, 4d and 5d elements, alloys and
  compounds'' in springermaterials.
\newblock Copyright 1986 Springer-Verlag Berlin Heidelberg.

\bibitem{Adachi1986}
K~Adachi, T~Uchiyama, M~Matsui, T~Miyazaki, M~Doi, and T~Miyazaki.
\newblock {Spin Glass of liquid-quenched Cu-Fe alloys}.
\newblock {\em J MAGN MAGN MATER}, pages 80--81, 1986.

\bibitem{Franz1973}
J.M. Franz and D.J. Sellmyer.
\newblock {Magnetic Interactions and High-Field Magnetization in Dilute
  Magnetic Alloys}.
\newblock {\em PHYS REV B}, 8(5), 1973.

\bibitem{Vedyaev1982}
AV~Vedyaev and V~Cherenkov.
\newblock Spin glass state in alloys of copper with manganese, iron and cobalt.
\newblock {\em SOV PHYS JETP-USSR}, 55(2):287--290, 1982.

\end{thebibliography}

\end{document}